\documentclass[twocolumn]{aastex63}

\newcommand\revision[1]{{{#1}}}

\usepackage{wasysym}

\received{December 16, 2020}
\accepted{February 10, 2021}
\submitjournal{PSJ}

\shorttitle{Reactivation of 259P/Garradd}
\shortauthors{Hsieh et al.}

\begin{document}

\title{The Reactivation of Main-Belt Comet 259P/Garradd (P/2008 R1)}

\correspondingauthor{Henry H.\ Hsieh}
\email{hhsieh@psi.edu}

\author[0000-0001-7225-9271]{Henry H.\ Hsieh}
\affiliation{Planetary Science Institute, 1700 East Fort Lowell Rd., Suite 106, Tucson, AZ 85719, USA}
\affiliation{Institute of Astronomy and Astrophysics, Academia Sinica, P.O.\ Box 23-141, Taipei 10617, Taiwan}

\author[0000-0002-7332-2479]{Masateru Ishiguro}
\affiliation{Department of Physics and Astronomy, Seoul National University, Gwanak, Seoul 151-742, Korea}

\author[0000-0003-2781-6897]{Matthew M.\ Knight}
\affiliation{Department of Physics, U.S. Naval Academy, 572C Holloway Rd., Annapolis, MD, 21402, USA}
\affiliation{Department of Astronomy, University of Maryland, 1113 Physical Sciences Complex, Building 415, College Park, MD 20742, USA}

\author[0000-0001-6765-6336]{Nicholas A.\ Moskovitz}
\affiliation{Lowell Observatory, 1400 W.\ Mars Hill Rd, Flagstaff, AZ 86011, USA}

\author[0000-0003-3145-8682]{Scott S.\ Sheppard}
\affiliation{Earth and Planets Laboratory, Carnegie Institution for Science, 5241 Broad Branch Road NW, Washington, DC 20015, USA}

\author[0000-0001-9859-0894]{Chadwick A.\ Trujillo}
\affiliation{Department of Physics and Astronomy, Northern Arizona University, Flagstaff, AZ 86011, USA}







\begin{abstract}
We present observations of main-belt comet 259P/Garradd from four months prior to its 2017 perihelion passage to five months after perihelion using the Gemini North and South telescopes.  The object was confirmed to be active during this period, placing it among seven MBCs confirmed to have recurrent activity.
We find \revision{an average} net \revision{pre-perihelion} dust production rate for 259P in 2017 of \revision{$\dot M_d=(4.6\pm0.2)$~kg~s$^{-1}$} \revision{(assuming grain densities of $\rho=2500$~kg~m$^{-3}$ and a mean effective particle size of ${\bar a_d}=2$~mm)} and a best-fit start date of \revision{detectable} activity of 2017~April~$22\pm1$, when the object was at a heliocentric distance of \revision{$r_h=(1.96\mp0.03)$~au} and a true anomaly of \revision{$\nu=(313.9\pm0.4)^{\circ}$}.  We estimate the effective active fraction of 259P's surface area to be from \revision{$f_{\rm act}\sim7\times10^{-3}$} to \revision{$f_{\rm act}\sim6\times10^{-2}$} (corresponding to effective active areas of \revision{$A_{\rm act}\sim8\times10^3$~m$^2$} to \revision{$A_{\rm act}\sim7\times10^4$~m$^2$}) at the start of its 2017 active period. A comparison of estimated total dust masses measured for 259P in 2008 and 2017 shows no evidence of changes in activity strength between the two active apparitions. The heliocentric distance of 259P's activity onset point is much smaller than those of other MBCs, suggesting that its ice reservoirs may be located at greater depths than on MBCs farther from the Sun, increasing the time needed for a \revision{solar irradiation-driven} thermal wave to reach subsurface ice.  We suggest that deeper ice on 259P could be a result of more rapid ice depletion caused by the object's closer proximity to the Sun compared to other MBCs.

\end{abstract}



\section{Introduction} \label{section:intro}

Comet 259P/Garradd (previously known as P/2008 R1) was discovered on 2008 September 2 \citep{garradd2008_259p} when it was at a heliocentric distance of $r_h=1.817$~au, a geocentric distance of $\Delta=0.938$~au, and a true anomaly of $\nu=18.3^{\circ}$.  With orbital elements placing it unambiguously in the main asteroid belt (semimajor axis of $a=2.727$~au, eccentricity of $e=0.342$, and inclination of $i=15.90^{\circ}$\revision{)}, it was quickly recognized as a member of the then-newly identified class of main-belt comets (MBCs).

MBCs orbit in the main asteroid belt, yet exhibit comet-like activity consistent with the sublimation of volatile material \citep{hsieh2006_mbcs,snodgrass2017_mbcs}.  They are a subset of the class of objects known as active asteroids, which \revision{exhibit comet-like mass loss that may be due to any of a number of mechanisms including sublimation, impacts, and rotational destabilization \citep{jewitt2015_actvasts_ast4}, but} have asteroidal orbits as defined by the Tisserand parameter with respect to Jupiter ($T_J$), \revision{which is given by}
\begin{equation}
    \revision{T_J = {a_J\over a} + 2\cos i\left[\left(1-e^2\right){a\over a_J}\right]^{1/2}}
\end{equation}
\revision{where $a_J=5.204$~au is the semimajor axis of Jupiter.  Small solar system bodies in the inner solar system are typically considered dynamically asteroidal if they have $T_J>3$, and dynamically cometary if they have $T_J<3$ \citep{kresak1979_cometasteroidinterrelations_ast1}, although sometimes a slightly higher $T_J$ threshold is used for classifying orbits as asteroidal \citep[e.g., $T_J>3.05$ or $T_J>3.08$;][]{tancredi2014_asteroidcometclassification,jewitt2015_actvasts_ast4} in order to account for slight real-world deviations from the idealized restricted three-body problem from which the $T_J$ parameter is derived. 259P has $T_J=3.217$, however, making it unambiguously dynamically asteroidal by this criterion. }

\setlength{\tabcolsep}{4.5pt}
\setlength{\extrarowheight}{0em}
\begin{table*}[htb!]
\caption{New 259P observations}
\centering
\smallskip
\footnotesize
\begin{tabular}{lccrcrccrccccr}
\hline\hline
\multicolumn{1}{c}{UT Date}
 & \multicolumn{1}{c}{Tel.$^a$}
 & \multicolumn{1}{c}{$N$$^b$}
 & \multicolumn{1}{c}{$t$$^c$}
 & \multicolumn{1}{c}{Filter}
 & \multicolumn{1}{c}{$\nu$$^d$}
 & \multicolumn{1}{c}{$r_h$$^e$}
 & \multicolumn{1}{c}{$\Delta$$^f$}
 & \multicolumn{1}{c}{$\alpha$$^g$}
 & \multicolumn{1}{c}{$m_{R,n}$$^h$}
 & \multicolumn{1}{c}{$m_{R,t}$$^i$}
 & \multicolumn{1}{c}{\revision{$H_{R,t}$$^j$}}
 & \multicolumn{1}{c}{$M_d$$^k$}
 & \multicolumn{1}{c}{$Af\rho$$^l$}
 \\
\hline
2013 Jan 25 & \multicolumn{4}{l}{Perihelion ...............................} & 0.0 & 1.798 & 2.758 &  5.5  & --- & --- & --- & \multicolumn{1}{c}{---} & \multicolumn{1}{c}{---} \\
2013 Aug 16 & Gemini-N & 10 &  1800 & $r'$ &  80.7 & 2.285 & 2.101 & 26.3 & 24.6$\pm$0.1 & --- & 19.6$\pm$0.1 & \revision{0.2$\pm$0.3} & 0.0$\pm$0.1 \\
2017 Mar 28 & Gemini-S &  1 &   300 & $r'$ & 304.4 & 2.032 & 1.850 & 29.3 & 23.9$\pm$0.2 & --- & 19.3$\pm$0.2 & \revision{0.9$\pm$0.6} & 0.2$\pm$0.1 \\ 
2017 Apr 26 & Gemini-S &  7 &   700 & $r'$ & 315.6 & 1.949 & 1.481 & 30.4 & 23.0$\pm$0.1 & --- & 18.9$\pm$0.1 & \revision{1.9$\pm$0.6} & 0.4$\pm$0.1 \\ 
2017 Apr 29 & Gemini-S &  7 &   700 & $r'$ & 316.8 & 1.942 & 1.446 & 30.3 & 22.9$\pm$0.1 & --- & 18.9$\pm$0.1 & \revision{2.1$\pm$0.6} & 0.5$\pm$0.1 \\ 
2017 Jun 30 & Gemini-S &  3 &   300 & $r'$ & 343.8 & 1.827 & 0.930 & 21.1 & 19.63$\pm$0.05 & 19.21$\pm$0.05 & 16.67$\pm$0.05 & \revision{28$\pm$3} & 6.6$\pm$0.7 \\ 
2017 Jul 01 & Gemini-S &  5 &   500 & $r'$ & 344.3 & 1.826 & 0.925 & 20.9 & 19.54$\pm$0.05 & 19.14$\pm$0.05 & 16.62$\pm$0.05 & \revision{30$\pm$3} & 7.1$\pm$0.7 \\ 
2017 Jul 18 & Gemini-S &  4 &   400 & $r'$ & 352.1 & 1.813 & 0.878 & 18.2 & 19.39$\pm$0.05 & 18.78$\pm$0.05 & 16.51$\pm$0.05 & \revision{33$\pm$3} & 6.8$\pm$0.7 \\ 
2017 Aug  4 & \multicolumn{4}{l}{Perihelion ...............................} & 0.0 & 1.809 & 0.877 & 18.4  & --- & --- & --- & \multicolumn{1}{c}{---} & \multicolumn{1}{c}{---} \\
2017 Sep 17 & Gemini-S &  1 &   300 & $r'$ &  20.5 & 1.838 & 1.073 & 26.8 & 19.46$\pm$0.05 & 18.65$\pm$0.05 & 15.54$\pm$0.05 & \revision{83$\pm$10} & 4.8$\pm$1.4 \\ 
2017 Nov 17 & Gemini-S &  2 &   600 & $r'$ &  46.5 & 1.964 & 1.684 & 30.2 & 20.68$\pm$0.05 & 20.14$\pm$0.05 & 15.77$\pm$0.05 & \revision{66$\pm$8} & 7.8$\pm$1.0 \\ 
2017 Dec 22 & Gemini-S &  4 &  1200 & $r'$ &  59.7 & 2.068 & 2.138 & 27.0 & 22.1$\pm$0.1 & 22.0$\pm$0.1 & 17.1$\pm$0.1 &  \revision{18$\pm$3} & 2.5$\pm$0.4 \\ 
2022 Feb 08 & \multicolumn{4}{l}{Perihelion ...............................} & 0.0 & 1.806 & 2.758 &  6.6  & --- & --- & --- & \multicolumn{1}{c}{---} & \multicolumn{1}{c}{---} \\
\hline
\hline
\multicolumn{14}{l}{$^a$ Telescope used.} \\
\multicolumn{14}{l}{$^b$ Number of exposures.} \\
\multicolumn{14}{l}{$^c$ Total integration time, in seconds.} \\
\multicolumn{14}{l}{$^d$ True anomaly, in degrees.} \\
\multicolumn{14}{l}{$^e$ Heliocentric distance, in au.} \\
\multicolumn{14}{l}{$^f$ Geocentric distance, in au.} \\
\multicolumn{14}{l}{$^g$ Solar phase angle (Sun-object-Earth), in degrees.} \\
\multicolumn{14}{l}{$^h$ Equivalent mean apparent $R$-band near-nucleus magnitude, measured within photometry apertures with radii of $4\farcs0$.} \\
\multicolumn{14}{l}{$^i$ Equivalent total mean apparent $R$-band magnitude, including the entire coma and tail, if present.} \\
\multicolumn{14}{l}{$^j$ Total absolute $R$-band magnitude, using $H,G$ phase function where $G=-0.08$.} \\
\multicolumn{14}{l}{$^k$ Estimated total dust mass, in $10^6$ kg, \revision{computed using Equation~\ref{equation:dust_mass} and} assuming $\rho_d\sim2500$~kg~m$^3$ \revision{and ${\bar a}_d=2$~mm}.} \\
\multicolumn{14}{l}{$^l$ $A(\alpha=0^{\circ})f\rho$ values, computed using \revision{Equation~\ref{equation:afrho} and} photometry apertures with radii of $4\farcs0$, in cm.} \\
\end{tabular}
\label{table:obs_259p}
\end{table*}

A series of observations conducted by \citet{jewitt2009_259p} shortly following 259P's discovery showed that its intrinsic brightness faded at a mean rate of approximately 0.01~mag~day$^{-1}$ over a period during which the comet ranged from $\nu=29.2^{\circ}$ to $\nu=48.6^{\circ}$.  Spectroscopic observations by the same authors using the Keck I telescope set an upper limit CN production rate of $Q_{\rm CN}=1.4\times10^{23}$~molecules~s$^{-1}$.  \citet{jewitt2009_259p} also noted that 259P is dynamically unstable on timescales of $\sim20-30$~Myr, suggesting that it may have been recently implanted in its current location from elsewhere in the solar system \citep[e.g.,][]{hsieh2016_tisserand}, unlike many other known MBCs that have been found to be dynamically stable on timescales of $>$100~Myr  and are therefore considered likely to have formed in situ \citep[e.g.,][]{hsieh2012_288p,hsieh2012_324p,hsieh2013_p2012t1}.  \citet{maclennan2012_259p} measured 259P's $R$-band absolute magnitude to be \revision{$H_R=(19.71\pm0.05)$~mag}, corresponding to an effective nucleus radius of \revision{$r_e=(0.30\pm0.02)$~km} (assuming an $R$-band albedo of $p_R=0.05$).  They also found a slope parameter value of $G_R=-0.08\pm0.05$, and an estimated photometric range (peak to trough) due to rotational variations of $\Delta m\sim0.6$~mag, although an actual rotation period could not be definitively identified.  259P is not associated with any currently known asteroid families \citep{hsieh2018_activeastfamilies}.

\citet{hsieh2017_259p} reported that observations of 259P showed it to be active again on 2017 April 26 and 29, during which time it ranged in heliocentric distance from $r_h=1.95$~au to $r_h=1.94$~au, and in true anomaly from $\nu=315^{\circ}$ to $\nu=317^{\circ}$.  These observations showing recurrent activity near perihelion after a period of inactivity away from perihelion \citep{maclennan2012_259p} are a strong indication that 259P's activity is due to sublimation of volatile material, rather than mechanisms such as collisional or rotational disruption which would not be expected to produce repeated activity near perihelion \citep[e.g.,][]{hsieh2012_scheila}, justifying its initial identification as a MBC by \citet{jewitt2009_259p}.

\section{Observations}\label{section:observations}

Observations of 259P were obtained on UT 2013 August 16 and several nights between 2017 March 28 and 2017 December 22 with the 8.1~m Gemini North (Gemini-N) telescope on Maunakea in Hawaii and the 8.1~m Gemini South (Gemini-S) telescope on Cerro Pach{\'o}n in Chile (Gemini Program IDs GN-2013A-Q-102, GS-2017A-LP-11, and GS-2017B-LP-11).  The 2013 observations were obtained approximately eight months after the object's 2013 perihelion passage (the object was unfortunately not observable closer to that perihelion passage), while the 2017 observations covered a period extending approximately four months prior to the object's 2017 perihelion passage to five months after that perihelion passage.

All observations on both Gemini telescopes were obtained using the telescopes' respective Gemini Multi-Object Spectrograph instruments \citep[GMOS;][]{hook2004_gmos,gimeno2016_gmoss} in imaging mode, Sloan $r'$-band filters, and  non-sidereal tracking.  Random dither offsets of up to $10''$ east or west, and north or south were applied to individual exposures.  All observations were conducted at airmasses of $<1.8$.  Standard bias subtraction, flat field correction, and cosmic ray removal were performed for all images using Python 3 code utilizing the {\tt ccdproc} package in Astropy \citep{astropy2018_astropy} and the {\tt L.A.Cosmic} python module\footnote{Written for python by Maltes Tewes (\url{https://obswww.unige.ch/\~tewes/cosmics\_dot\_py/})} \citep{vandokkum2001_lacosmic,vandokkum2012_lacosmic}.
To maximize signal-to-noise ratios (S/N), we constructed composite images of the object for each night of data by shifting and aligning individual images on the object's photocenter using linear interpolation and then adding them together. 

Photometry measurements of 259P and background reference stars were performed using Image Reduction and Analysis Facility software \citep[IRAF;][]{tody1986_iraf,tody1993_iraf}, with absolute photometric calibration performed using field star magnitudes from the {\tt refcat} all-sky stellar reference catalog \citep{tonry2018_refcat}.  Conversion of $r'$-band Gemini and Pan-STARRS1 photometry (used in the {\tt refcat} catalog) to $R$-band was accomplished using transformations derived by \citep{tonry2012_ps1} and by R.\ Lupton ({\tt http://www.sdss.org/}).  At least five reference stars were used per image when possible, although due to the much greater image depth of our images compared to the underlying data used to construct the {\tt refcat} catalog, sometimes only one or two stars in a field were both unsaturated in our data and found in the {\tt refcat} catalog, and thus available to be used for photometric calibration.

For most of our data, field star photometry was performed using circular apertures with sizes chosen using curve-of-growth analyses of each night of data, where background statistics were measured in an annulus around each star with a radius chosen to be large enough to avoid most of the flux from the star.  However, for our 2017 March 28 data, the relatively dense star field and considerable amount of field-star trailing necessitated a modified approach to field star photometry.  For these data, each image was first rotated using the IRAF {\tt rotate} task such that star trails were horizontal in the image, and field star photometry was then performed using rectangular apertures that enclosed star trails that were judged to be relatively isolated. Background statistics of nearby regions of blank sky were measured using manually defined rectangular apertures.

Meanwhile, target photometry when 259P was inactive or only minimally active, and field stars were only minimally trailed (on 2013 August 16, and 2017 March 28 through 2017 April 29) was performed using circular apertures with sizes chosen using curve-of-growth analyses of each night of data.  For these data, background statistics were measured in nearby but non-adjacent regions of blank sky to avoid potential dust contamination from the object or nearby field stars.  When 259P exhibited stronger activity (2017 June 30 to 2017 December 22), we measured the total flux from each object in our composite images from each night using rectangular photometry apertures with sizes and orientations chosen to enclose as much of the visible dust cloud as possible without introducing significant field star contamination.  On these nights, we also performed near-nucleus photometry using circular apertures with $4\farcs0$ radii.  In both cases, background sky levels were measured from nearby but non-adjacent regions of blank sky and subtracted to obtain net fluxes.  \citet{jewitt2009_259p} only reported photometry using circular apertures \revision{with $2\farcs2$ radii} that did not fully enclose the object's visible dust cloud in their 2008 observations. \revision{As such,} we used \revision{the same methods described here for measuring near-nucleus and total brightnesses for data when 259P was visibly active} to reanalyze their only publicly available \revision{2008} observations \revision{in order to provide a better basis for comparison with our 2017 data}.  \revision{Those data} were originally obtained on 2008 October 22 using the Faint Object Camera And Spectrograph instrument \citep[FOCAS;][]{yoshida2000_focas,kashikawa2002_focas} at Subaru Observatory and \revision{retrieved} from Subaru's Subaru-Mitaka-Okayama-Kiso Archive (SMOKA) system \citep{baba2002_smoka}.

\begin{figure}[tbp]
\centerline{\includegraphics[width=3.3in]{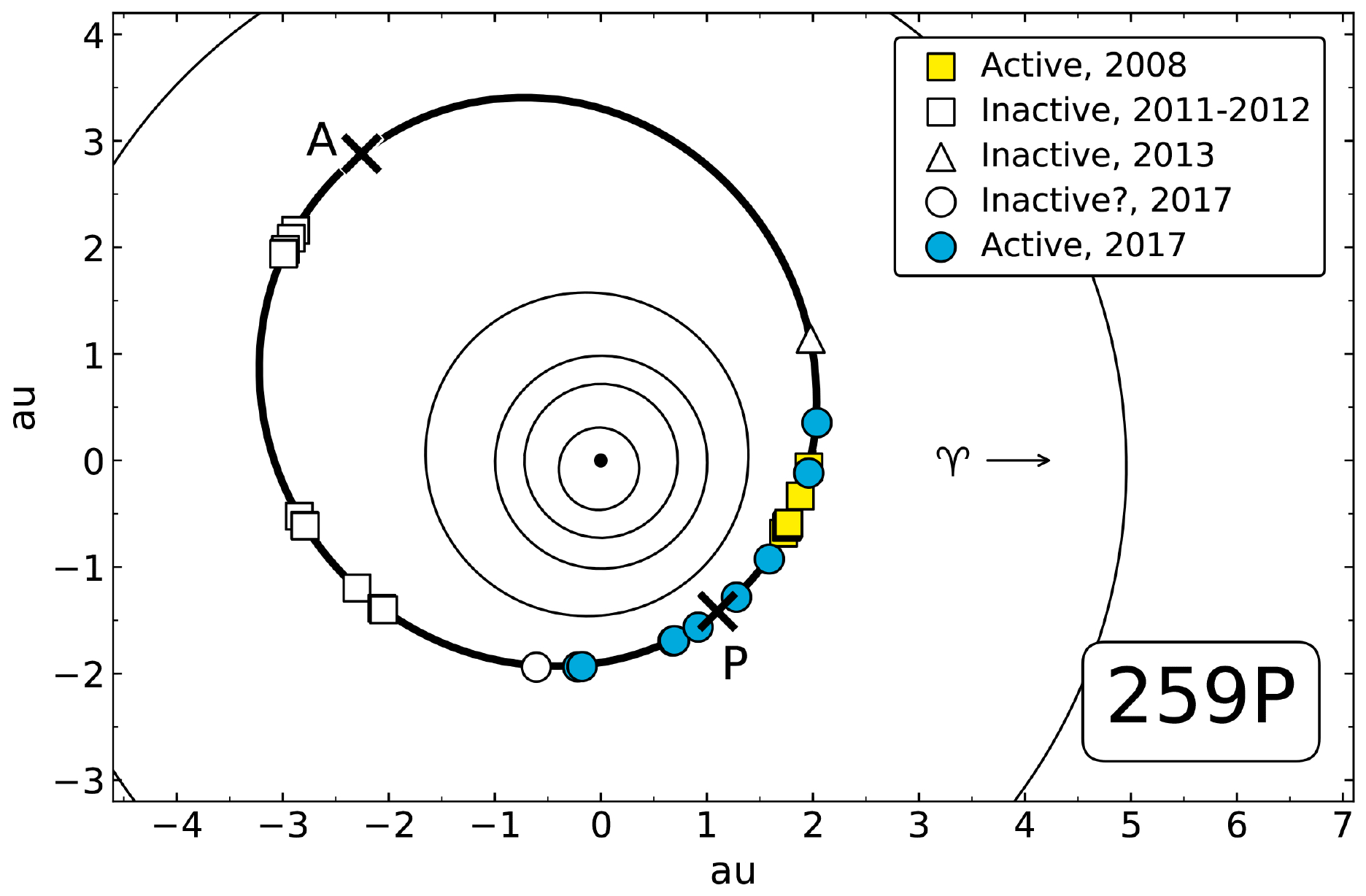}}
\caption{\small Orbit position plot with the Sun (black dot) at the center, the orbits of Mercury, Venus, Earth, Mars, and Jupiter shown as thin black lines, and the orbit of 259P shown as a thick black line. 259P's perihelion (P) and aphelion (A) are marked with X's.  Open and filled square, triangle, and circle symbols mark positions of 259P when it was observed to be either active or inactive, as indicated by the figure legend. An arrow and the $\aries$ symbol indicate the direction of the vernal equinox (ecliptic longitude of 0$^{\circ}$).
}
\label{figure:orbit_259p}
\end{figure}
\begin{figure}[tbp]
\centerline{\includegraphics[width=3.3in]{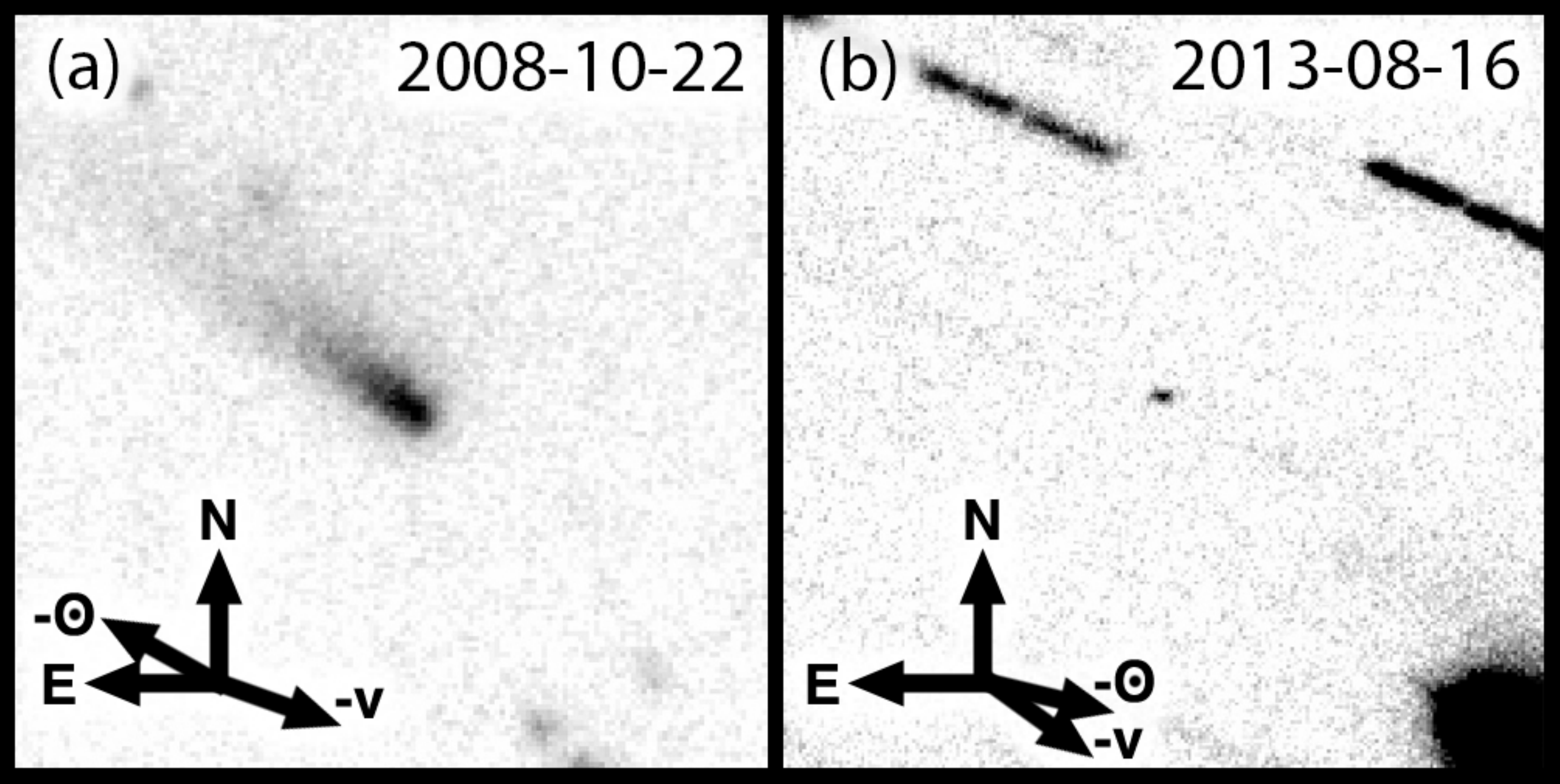}}
\caption{\small Composite (a) $R$-band and (b) $r'$-band images of 259P (at the center of each panel) constructed from data listed in Table~\ref{table:obs_259p}, where the object is active in panel (a) and we have determined that the object is likely to be inactive at the time of the observations in panel (b).  Both panels are $30''\times30''$ in size, with north (N), east (E), the antisolar direction ($-\odot$), and the negative heliocentric velocity vector ($-v$), as projected on the sky, marked.  Observation dates in YYYY-MM-DD format are listed in the upper right corner of each panel.}
\label{figure:images_259p_2008_2013}
\end{figure}
\begin{figure*}[tbp]
\centerline{\includegraphics[width=7.0in]{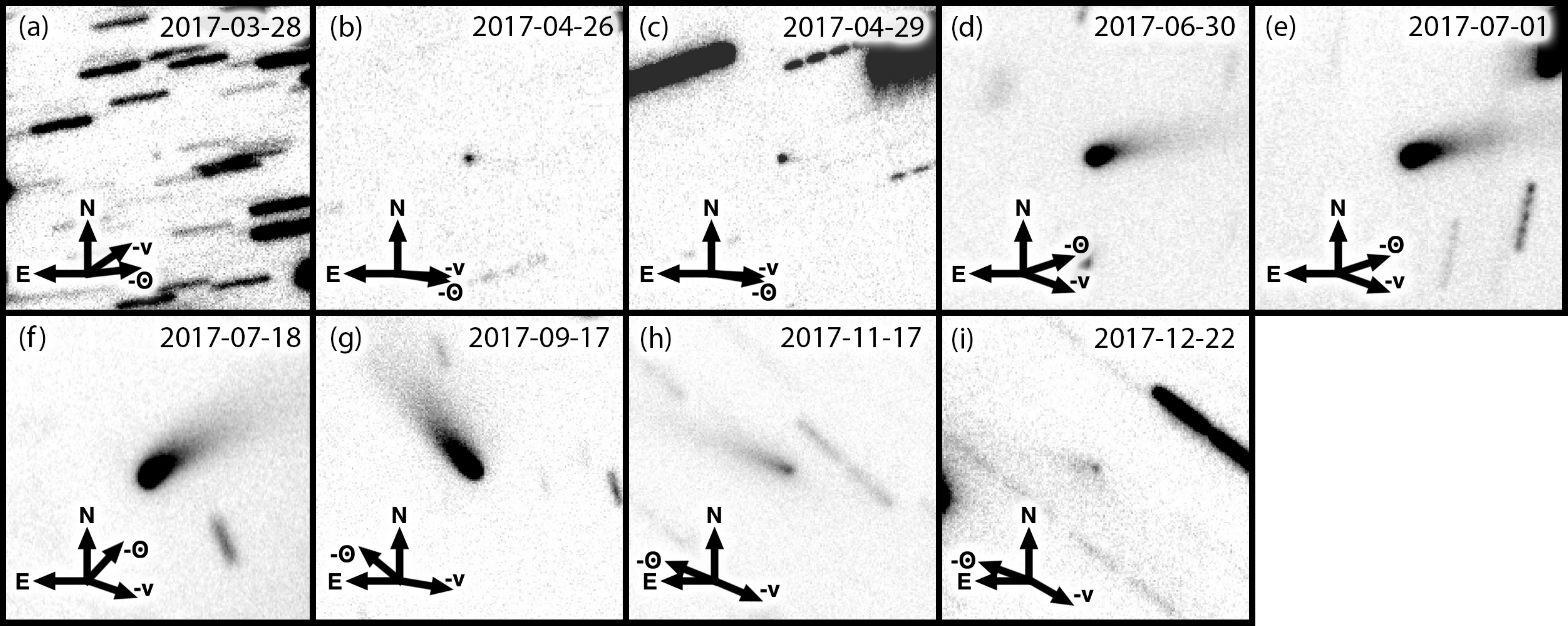}}
\caption{\small Composite $r'$-band images of 259P (at the center of each panel) constructed from data listed in Table~\ref{table:obs_259p}, where we have determined that the object is likely to be inactive at the time of the observations in panel (a), and active in panels (b) through (j).  All panels are $30''\times30''$ in size, with north (N), east (E), the antisolar direction ($-\odot$), and the negative heliocentric velocity vector ($-v$), as projected on the sky, marked.  Observation dates (in YYYY-MM-DD format) are listed in the upper right corner of each panel.}
\label{figure:images_259p}
\end{figure*}

Details of our observations of 259P are listed in Table~\ref{table:obs_259p}. We also mark the orbit positions of both the observations reported here and observations previously reported in \citet{jewitt2009_259p} and \citet{maclennan2012_259p} in Figure~\ref{figure:orbit_259p}.  Composite images of the object during each night of observations reported in this work are shown in Figures~\ref{figure:images_259p_2008_2013} and \ref{figure:images_259p}.

\setlength{\tabcolsep}{4.0pt}
\setlength{\extrarowheight}{0em}
\begin{table*}[htb!]
\caption{Previously reported 259P observations}
\centering
\smallskip
\footnotesize
\begin{tabular}{lcrccrccccccc}
\hline\hline
\multicolumn{1}{c}{UT Date}
 & \multicolumn{1}{c}{Tel.$^a$}
 & \multicolumn{1}{c}{$\nu$$^b$}
 & \multicolumn{1}{c}{$r_h$$^c$}
 & \multicolumn{1}{c}{$\Delta$$^d$}
 & \multicolumn{1}{c}{$\alpha$$^e$}
 & \multicolumn{1}{c}{$m_{R,n}$$^f$}
 & \multicolumn{1}{c}{$m_{R,t}$$^g$}
 & \multicolumn{1}{c}{$H_{R,t}$$^h$}
 & \multicolumn{1}{c}{$H'_{R,t}$$^{i}$}
 & \multicolumn{1}{c}{$M_d$$^j$}
 & \multicolumn{1}{c}{$M'_d$$^k$}
 & \multicolumn{1}{c}{\revision{$Af\rho$$^l$}}
 \\
\hline
\multicolumn{10}{l}{{\it Previously reported photometry$^m$}} \\
~2008 Sep 26 & UH &  29.2 & 1.853 & 1.090 & 26.4 & 19.69$\pm$0.05 & 18.1$\pm$0.1 & $<$16.5 & 15.0$\pm$0.1 & $>$15.8 & \revision{142$\pm$16} & \revision{17.4$\pm$2.1} \\
~2008 Sep 30 & Keck &  30.9 & 1.861 & 1.122 & 27.1 & 19.88$\pm$0.03 & 18.3$\pm$0.1 & $<$16.6 & 15.1$\pm$0.1 & $>$14.5 & \revision{130$\pm$14} & \revision{15.5$\pm$1.8} \\
~2008 Sep 30 & UH &  30.9 & 1.861 & 1.122 & 27.1 & 19.93$\pm$0.03 & 18.4$\pm$0.1 & $<$16.7 & 15.1$\pm$0.1 & $>$13.8 & \revision{124$\pm$14} & \revision{14.8$\pm$1.7} \\
~2008 Oct 01 & UH &  31.4 & 1.863 & 1.131 & 27.2 & 19.86$\pm$0.02 & 18.3$\pm$0.1 & $<$16.6 & 15.0$\pm$0.1 & $>$15.2 & \revision{136$\pm$15} & \revision{16.1$\pm$1.8} \\
~2008 Oct 02 & UH &  31.8 & 1.865 & 1.139 & 27.4 & 19.90$\pm$0.03 & 18.3$\pm$0.1 & $<$16.6 & 15.0$\pm$0.1 & $>$14.9 & \revision{134$\pm$15} & \revision{15.7$\pm$1.8} \\
~2008 Oct 03 & UH &  32.3 & 1.867 & 1.148 & 27.5 & 20.17$\pm$0.06 & 18.6$\pm$0.1 & $<$16.9 & 15.3$\pm$0.1 & $>$11.7 & \revision{106$\pm$13} & \revision{12.2$\pm$1.6} \\
~2008 Oct 22 & Subaru &  40.4 & 1.909 & 1.331 & 29.4 & 21.09$\pm$0.04 & 19.5$\pm$0.1 & $<$17.3 & 15.7$\pm$0.1 &  $>$7.3 & \revision{68$\pm$8} &  \revision{6.5$\pm$0.9} \\
~2008 Nov 11 & UH &  48.6 & 1.962 & 1.560 & 30.0 & 21.21$\pm$0.10 & 19.6$\pm$0.1 & $<$17.0 & 15.4$\pm$0.1 &  $>$9.9 & \revision{91$\pm$14} &  \revision{7.6$\pm$1.2} \\
\hline
\multicolumn{10}{l}{{\it Re-measured photometry$^n$}} \\
~2008 Oct 22 & Subaru &  40.4 & 1.909 & 1.331 & 29.4 & 20.26$\pm$0.05 & 19.51$\pm$0.05 & 15.7$\pm$0.1 & --- & \revision{68$\pm$9} & --- & \revision{8.2$\pm$1.1} \\
\hline
\hline
\multicolumn{13}{l}{$^a$ Telescope (UH: University of Hawaii 2.2~m telescope; Keck: 10~m Keck I telescope; Subaru: 8.2~m Subaru Observatory)} \\
\multicolumn{13}{l}{$^b$ True anomaly, in degrees.} \\
\multicolumn{13}{l}{$^c$ Heliocentric distance, in au.} \\
\multicolumn{13}{l}{$^d$ Geocentric distance, in au.} \\
\multicolumn{13}{l}{$^e$ Solar phase angle (Sun-object-Earth), in degrees.} \\
\multicolumn{13}{l}{$^f$ Equivalent mean apparent $R$-band nucleus magnitude, measured within a circular photometry aperture, as reported by } \\
\multicolumn{13}{l}{$~~~~$ \citet{jewitt2009_259p} for previously reported photometry, or measured as part of this work for re-measured photometry.} \\
\multicolumn{13}{l}{$^g$ Equivalent total mean apparent $R$-band magnitude, including the entire coma and tail, if present, as estimated from scaling } \\
\multicolumn{13}{l}{$~~~~$ previously reported photometry \revision{(listed in this table as $m_{R,n}$)} as described in Section~\ref{subsection:photometric_analysis}, or measured as part of this work,} \\
\multicolumn{13}{l}{$~~~~$ as applicable. } \\
\multicolumn{13}{l}{$^h$ Total absolute $R$-band magnitude or upper limit based on previously reported photometry, as applicable, using an $H,G$} \\
\multicolumn{13}{l}{$~~~~$ phase function where $G=-0.08$.} \\
\multicolumn{13}{l}{$^i$ Total absolute $R$-band magnitude, using an $H,G$ phase function where $G=-0.08$, \revision{as} estimated from \revision{scaling} previously} \\
\multicolumn{13}{l}{$~~~~$  reported photometry \revision{(listed in this table as $m_{R,n}$)} as described in Section~\ref{subsection:photometric_analysis}} \\
\multicolumn{13}{l}{$^j$ Estimated total dust mass or lower limit based on previously reported photometry, as applicable, in $10^6$ kg, } \\
\multicolumn{13}{l}{$~~~~$ assuming $\rho_d\sim2500$~kg~m$^{-3}$ \revision{and ${\bar a}_d=2$~mm}.} \\
\multicolumn{13}{l}{$^k$ Estimated total dust mass, in $10^6$ kg, assuming $\rho_d\sim2500$~kg~m$^{-3}$ \revision{and ${\bar a}_d=2$~mm}, \revision{as estimated from scaling} } \\
\multicolumn{13}{l}{$~~~~$ previously reported photometry \revision{(listed in this table as $m_{R,n}$)} as described in Section~\ref{subsection:photometric_analysis}} \\
\multicolumn{13}{l}{\revision{$^l$ $A(\alpha=0^{\circ})f\rho$ values, computed using Equation~\ref{equation:afrho} and photometry apertures with radii of $2\farcs2$, in cm.}} \\
\multicolumn{13}{l}{$^m$ Computations based on photometry reported by \citet{jewitt2009_259p}.} \\
\multicolumn{13}{l}{$^n$ Computations based on photometry performed as part of this work, as described in Section~\ref{section:observations}.} \\
\end{tabular}
\label{table:obs_259p_previous}
\end{table*}

\section{Results and Analysis}\label{section:results}

\subsection{Photometric Analysis}\label{subsection:photometric_analysis}

\begin{figure*}[tbp]
\centerline{\includegraphics[width=4.0in]{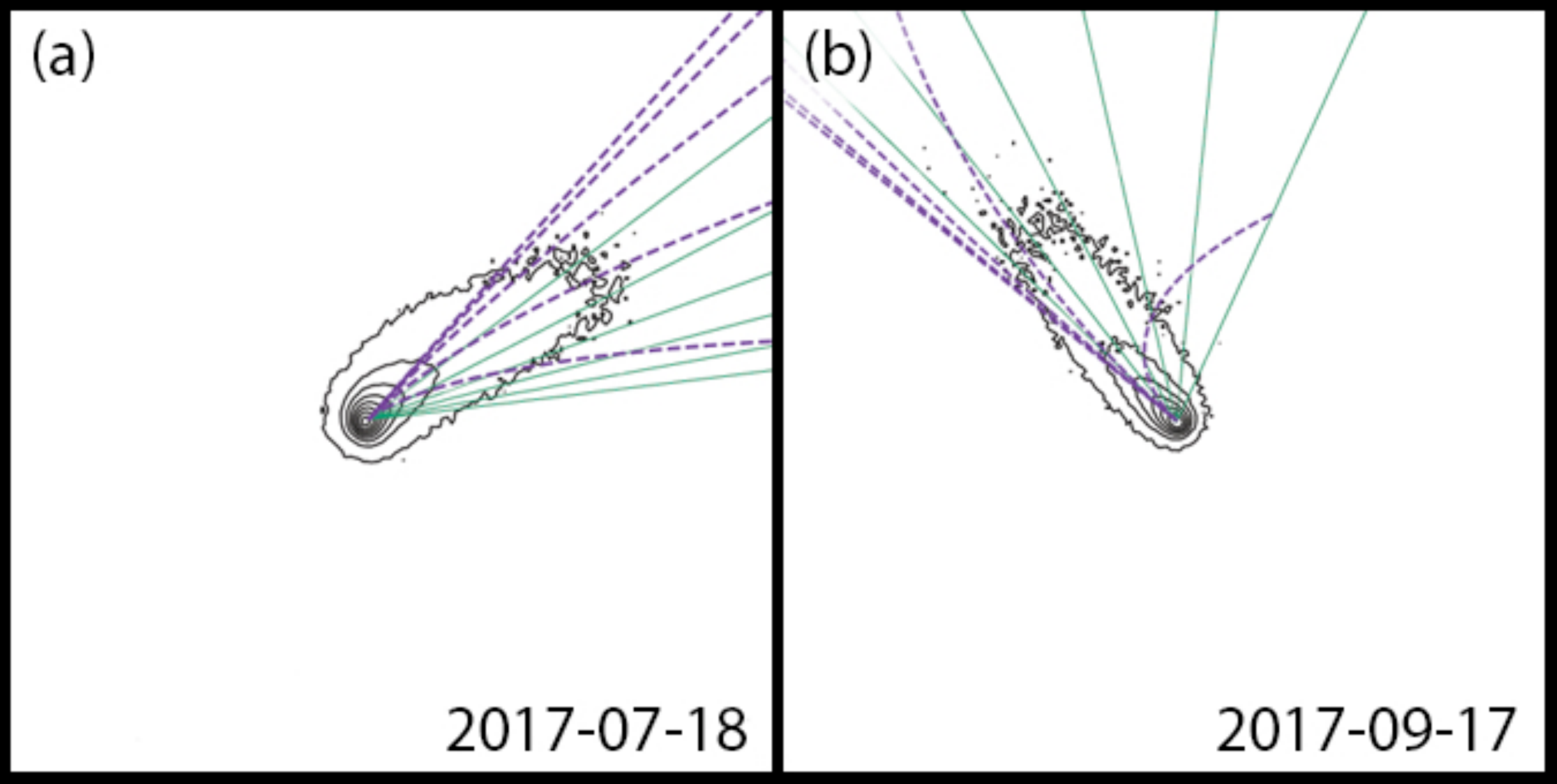}}
\caption{\small Syndynes (dashed purple lines) and synchrones (solid green lines) for 259P on (a) 2017 July 18 and (b) 2017 September 17, overlaid on \revision{log-spaced contour plots corresponding to} image data for those dates \revision{(plotted at the same spatial scale as in Figure~\ref{figure:images_259p})}. Syndynes from top to bottom in panel (a) and from left to right in panel (b) approximately correspond to particle sizes of \revision{1~cm,} 1~mm, 100~$\mu$m, 10~$\mu$m, and 1~$\mu$m, respectively.  Synchrones from top to bottom in panel (a) and from left to right in panel (b) correspond to particles ejected 15, 30, 45, 60, 75, and 90 days prior to the observation date.  }
\label{figure:syndynes_synchrones}
\end{figure*}

\begin{figure*}[tbp]
\centerline{\includegraphics[width=7.0in]{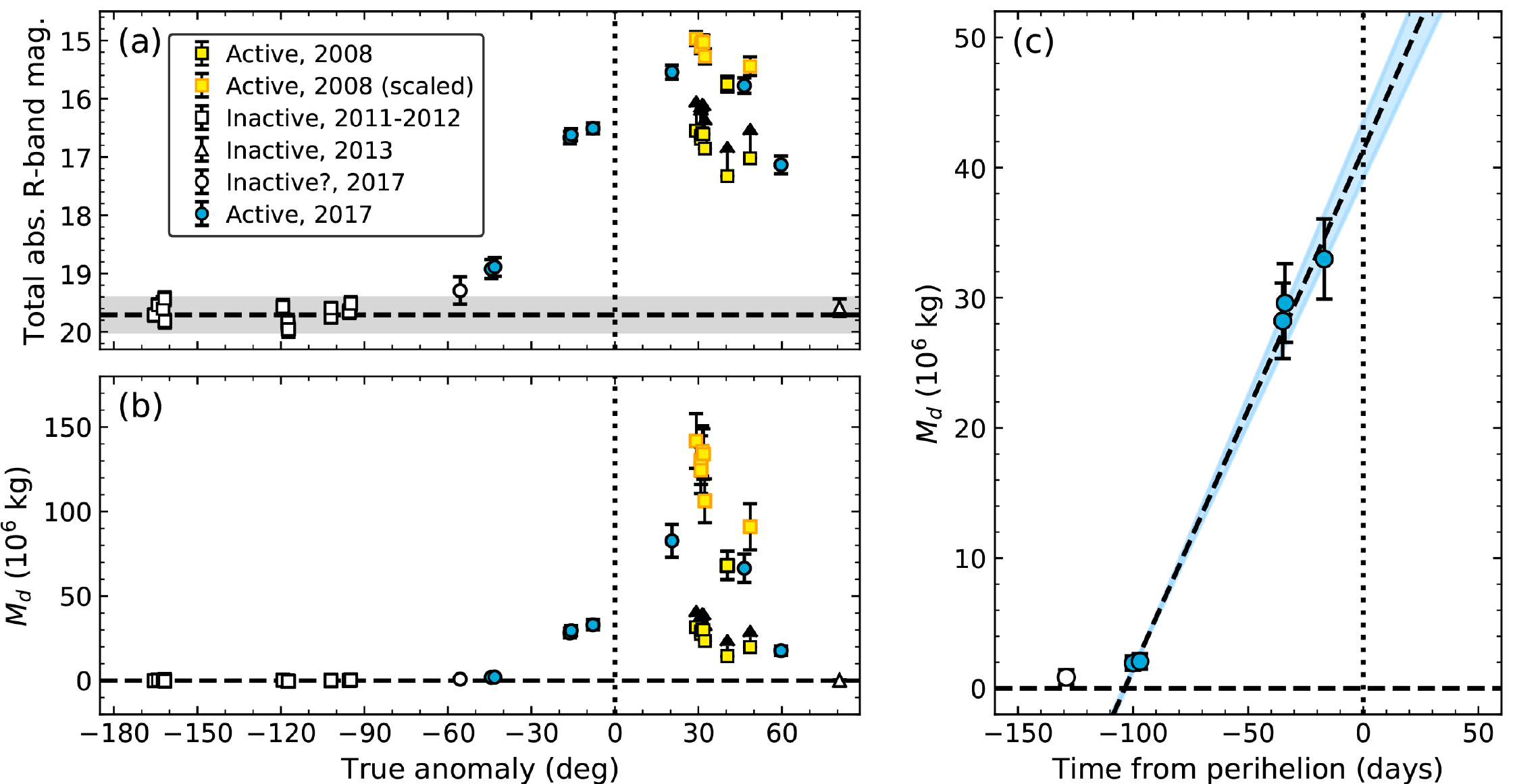}}
\caption{\small (a) Plot of total absolute $R$-band magnitudes (or upper limits, if marked with arrows instead of error bars) of 259P during various active and inactive periods from 2008 to 2017 as a function of true anomaly ($\nu$).  The expected magnitude of the inactive nucleus is marked with a horizontal dashed black line with the estimated range of potential magnitude variation due to the object's rotational lightcurve marked with a shaded gray region, while perihelion is marked with a dotted vertical line. (b) Total estimated dust masses (or lower limits, if marked with arrows instead of error bars) for 259P during the same periods of observations as in panel (a) plotted as a function of $\nu$.  The excess dust mass expected for the inactive nucleus (i.e., zero) is marked with a horizontal dashed black line, while perihelion is marked with a dotted vertical line.  (c) Estimated total dust masses for 259P during observations from 2017 plotted versus time from perihelion (where negative values denote time before perihelion and positive values denote time after perihelion).  A diagonal dashed line shows a linear fit to data obtained between 2017 April 26 and 2017 July 18 ($-55.6^{\circ} < \nu < -7.9^{\circ}$), reflecting an estimate of the average net dust production rate over this period and allowing us to estimate the onset time of activity, while the shaded blue region shows the 1-$\sigma$ range of uncertainty of the linear fit.  In all panels, data corresponding to 259P's 2008 active period are marked with yellow-filled squares, data corresponding to the 2011-2012 inactive period are marked with open squares, and data corresponding to the 2017 active period are marked with blue-filled circles.  Data corresponding to observations in 2013 and 2017 when the object was determined likely to be inactive are marked with an open triangle and an open circle, respectively.  Total absolute $R$-band magnitudes and excess dust masses estimated from previously reported photometry as described in Section~\ref{subsection:photometric_analysis} are marked with yellow-filled squares with orange outlines.}
\label{figure:excess_dust}
\end{figure*}


In order to quantitatively characterize 259P's 2017 activity evolution and compare it to those of other MBCs, we use photometric measurements made of 259P in this work to estimate the amounts of excess dust present in those observations.
First, following \citet{jewitt2014_133p} and assuming that the maximum dust particle radius, $a_{d,{\rm max}}$, is much larger than the minimum dust particle radius, $a_{d,{\rm min}}$ (i.e., $a_{d,{\rm max}}\gg a_{d,{\rm min}}$), we can compute an approximate mean effective particle radius (by mass), ${\bar a}_d$, weighted by size distribution, scattering cross-section, and residence time, using
\begin{equation}
    \revision{{\bar a}_d \sim {a_{d,{\rm max}}\over5}}
    \label{equation:effective_particle_size1}
\end{equation}
\revision{which applies to power law distributions with $q=3.25$.}

Size distributions and ranges of dust grains have not yet been reported for 259P.  As such, for the purposes of our dust mass estimates, we refer to a dust modeling study performed for another MBC, P/2015 X6 (PANSTARRS), whose semimajor axis ($a=2.755$~au) is similar to that of 259P ($a=2.727$~au). In that analysis, \citet{moreno2016_p2015x6} found \revision{a power-law distribution of particles with} minimum and maximum dust grain radii of $a_{d,{\rm min}}=1$~$\mu$m and $a_{d,{\rm max}}=1$~cm\revision{, respectively, and a power law index of $-3.3$ (which is close to the power law index of $q=3.25$ assumed for Equation~\ref{equation:effective_particle_size1}) to be consistent with their observations}.  \revision{Overlaying a syndyne-synchrone grid} \citep[][]{finson1968_cometdustmodeling1} \revision{on our data for comparison}, we find these values to be reasonably plausible for 259P's activity (Figure~\ref{figure:syndynes_synchrones}).
Thus, adopting \revision{1~cm for the value of $a_{d,{\rm max}}$} in Equation~\ref{equation:effective_particle_size1}, we compute a mean effective particle radius of \revision{${\bar a}_d\sim2$~mm}.

Using this mean effective particle radius and the phase function derived by \citet{maclennan2012_259p}\revision{, and assuming that ejected dust exhibits the same phase darkening behavior as the nucleus}, we can estimate the amounts of excess dust present in the 2017 observations of 259P reported in this work (Section~\ref{section:observations}). Following \citet{hsieh2014_324p}, we estimate the total mass, $M_d$, of visible ejected dust using
\begin{equation}
    \revision{M_d = {4\over 3}\pi r^2_N {\bar a_d}\rho_d \left(A_d\over A_N\right),}
\end{equation}
\revision{where the ratio of the total scattering surface area of dust to that of the nucleus, $A_d/A_N$, is given by}
\begin{equation}
    \revision{{A_d\over A_N} = \left({1-10^{0.4(H_{R,t}-H_R)} \over 10^{0.4(H_{R,t}-H_R)} }\right),}
\end{equation}
\revision{giving}
\begin{equation}
M_d = {4\over 3}\pi r^2_N {\bar a_d}\rho_d \left({1-10^{0.4(H_{R,t}-H_R)} \over 10^{0.4(H_{R,t}-H_R)} }\right)
\label{equation:dust_mass}
\end{equation}
where \revision{$H_{R,t}$} is the equivalent total absolute magnitude of the active nucleus at $r_h=\Delta=1$~au and $\alpha=0^{\circ}$ computed using the $H,G$ phase function and the best-fit $G$ parameter determined by \citet{maclennan2012_259p}.  We assume dust grain densities of $\rho_d$$\,=\,$2500~kg~m$^{-3}$, consistent with CI and CM carbonaceous chondrites, which are associated with primitive C-type objects like the MBCs \citep{britt2002_astdensities_ast3}. \revision{We note that dust masses and certain other parameters derived from computed dust masses (dust production rates, and active areas and fractions) vary proportionally with respect to assumed average grain density and mean particle size (see Equations~\ref{equation:dust_mass}, \ref{equation:active_area}, \ref{equation:active_fraction}), and so different estimates or assumptions for these values could produce results that differ by a factor of a few or more from those reported here.  As such, in the event that improved constraints on the average grain density or mean particle size become available in the future, the values of these derived quantities can and should be revised accordingly.}

For reference, we also \revision{use our measured near-nucleus photometry to} compute $A(\alpha=0^{\circ})f\rho$ values \citep[hereafter, $Af\rho$;][]{ahearn1984_bowell}, given by
\begin{equation}
A\revision{(\alpha=0^{\circ})}f\rho = {(2r_h\Delta)^2\over \rho}10^{0.4[m_\odot-m_{R,d}(r_h,\Delta,0)]}
\label{equation:afrho}
\end{equation}
where $r_h$ is in au, $\Delta$ is in cm, $\rho$ is the physical radius in cm of the photometry aperture used to measure the magnitude of the comet at the distance of the comet, and $m_{R,d}(r_h,\Delta,0)$ is the phase-angle-\revision{normalized} (to $\alpha=0^{\circ}$) $R$-band magnitude of the excess dust mass of the comet (i.e., with the flux contribution of the nucleus subtracted from the measured total \revision{flux}).  We note, however, that this parameter is not always a reliable measurement of the dust contribution to comet photometry in cases of non-spherically symmetric comae \citep[e.g.,][]{fink2012_afrho}.

The results of all calculations described above are shown in Table~\ref{table:obs_259p} for observations of 259P reported in this work.
\revision{For observations obtained when 259P did not have a visibly extended appearance (i.e., from 2013 August 16 to 2017 April 29) and  therefore a separate procedure was not required to measure the comet's total brightness, total absolute magnitudes and excess dust masses are computed directly from near-nucleus photometry measurements. In these cases, listed uncertainties do not include the potential photometric variability of the nucleus due to rotation.}
We also plot total absolute magnitudes and computed excess dust masses as functions of $\nu$ and days relative to perihelion in Figure~\ref{figure:excess_dust}.

For later comparison (see Section~\ref{subsection:previous_activity_analysis}), we also compute estimated excess dust masses for observations previously reported by \citet{jewitt2009_259p}.  Specifically, we compute the excess dust mass present on 2008 October 22 with the same methods as we use for our 2017 data \revision{(i.e., using a rectangular photometry aperture whose size and orientation is chosen to enclose as much of the visible dust cloud as possible while attempting to avoid significant field star contamination)}, using the publicly available data from Subaru from that date (see Section~\ref{section:observations}), and also use \revision{Equation~\ref{equation:dust_mass}} to derive lower limit excess dust masses from the reported near-nucleus photometry (assumed to give upper limits to the total magnitudes of the comet at the times of observation).

In order to compare similar quantities from 259P's 2008 and 2017 active apparitions, however, we wish to estimate the total dust masses for 259P in 2008, instead of \revision{just lower limits} from near-nucleus photometry.  To do this, we compare the reported near-nucleus magnitude of the comet on 2008 October 22 to the total magnitude we measure for the same data (obtained from the Subaru archive), where we find a difference of 1.58~mag between the two measurements.  Assuming an approximately constant ratio of near-nucleus fluxes measured by \citet{jewitt2009_259p} and the total fluxes that would be measured with our methods, we apply the same magnitude offset to all photometry from 2008 to derive adjusted total magnitudes.  We then use these adjusted total magnitudes to compute approximate total excess dust masses (Table~\ref{table:obs_259p_previous}), and plot our results in Figure~\ref{figure:excess_dust}.  The ratio of near-nucleus flux and total flux for the comet is of course not expected to remain exactly constant for all 2008 observations, and so the dust masses estimated using the method described here should not be regarded as very accurate.  Nonetheless, they provide at least an approximate basis upon which we can compare the strengths of 259P's activity in 2008 and in 2017.

To estimate 259P's \revision{average} net dust production rate, $\dot M_d$, \revision{over the period of time spanned by our pre-perihelion observations (during which the object's heliocentric distance changes by just 0.136~au, from $r_h=1.949$~au to $r_h=1.813$~au)}, and also \revision{estimate} the approximate onset time of the observed activity, we fit a linear function to excess dust masses computed from data obtained between 2017 April 26 and 2017 July 18, when measured excess dust masses appear to increase approximately linearly.  Following the calculations detailed by \citet{hsieh2015_ps1mbcs}, we find that the heliocentric distance change over this period
corresponds to a $\sim$19\%$-$38\% increase in the water sublimation rate on the object's surface \revision{and an increase in the equilibrium surface temperature of $\sim$1$-$2~K}, depending on whether a subsolar or isothermal approximation is assumed.  The resulting \revision{average net} dust production rate and corresponding activity start date we find are of course subject to numerous sources of uncertainty including \revision{variations in} the \revision{real} dust production rate as a function of time and the unknown rotational phases of the object at the times when each photometric point was obtained.  \revision{We also emphasize that the calculated average dust production rate is a net production rate and not an absolute production rate, given that the rate of dissipation and removal of dust from the visible dust tail by solar radiation pressure over this period is unconstrained and thus not considered as part of this analysis. }

We find a best-fit \revision{average} net dust production rate of $\dot M_d = \revision{(4.6\pm0.3)}$~kg~s$^{-1}$ over the time period under consideration and a best-fit start date for \revision{detectable} activity of 2017 April 22$\pm$1 ($\sim$104 days prior to perihelion), when the object was at $r_h=\revision{(1.96\mp0.03)}$~au and $\nu=\revision{(313.9\pm0.4)}^{\circ}$ (where we mark the uncertainty in $r_h$ using $\mp$ to indicate that a positive error in the best-fit start date corresponds to a smaller heliocentric distance, and vice versa, due to the object being inbound at the time in question).  Assuming \revision{a time-averaged} dust-to-gas ratio (by mass) of $f_{dg}=5$ \citep[cf.][]{hsieh2018_238p288p}, this computed dust production rate corresponds to a water production rate of \revision{$Q_{\rm H_2O}\sim3\times10^{25}$~molecules~s$^{-1}$} (assuming water to be the dominant volatile material, where the mass of one water molecule is $m_{\rm molecule}\sim3\times10^{-26}$~kg).  We note that this inferred water production rate is smaller than the upper limit water production rate of $Q_{\rm H_2O}<5\times10^{25}$~molecules~s$^{-1}$ derived from spectroscopic observations of 259P when it was active in 2008 \citep{jewitt2009_259p}.  This suggests that those observations may not have been sensitive enough to detect the level of outgassing present, as is suspected for most such attempts to date at detecting sublimation products from MBCs \citep[e.g.,][]{snodgrass2017_mbcs}, and not that outgassing was necessarily absent.

At the midpoint of the time period covered by this fitting analysis, water production rates are expected to range from $\dot m_w\sim1.3\times10^{-5}$~kg~s$^{-1}$~m$^{-2}$ in the isothermal (or ``fast rotator'') approximation to $\dot m_w\sim1.1\times10^{-4}$~kg~s$^{-1}$~m$^{-2}$ in the subsolar (or ``flat slab'') approximation for a sublimating graybody in \revision{local} thermal equilibrium.  Assuming the nucleus to be a spherical body with an effective radius of $r_N=300$~m \citep{maclennan2012_259p}, we can use
\begin{equation}
A_{\rm act} = {{\dot M}_d\over  f_{dg}{\dot m}_w}
\label{equation:active_area}
\end{equation}
and
\begin{equation}
f_{\rm act} = {A_{\rm act}\over 4\pi r_N^2}
\label{equation:active_fraction}
\end{equation}
to determine the effective active area, $A_{\rm act}$, and effective active fraction, $f_{\rm act}$ of the object's surface, respectively.  Assuming $f_{dg}=5$ as before, we find estimated effective active areas and active fractions ranging from \revision{$A_{\rm act}\sim8\times10^3$~m$^2$ and $f_{\rm act}\sim7\times10^{-3}$} (in the subsolar approximation) to \revision{$A_{\rm act}\sim7\times10^4$~m$^2$ and $f_{\rm act}\sim6\times10^{-2}$} (in the isothermal approximation) at the start of 259P's 2017 active period.


\subsection{Comparison with previous activity}\label{subsection:previous_activity_analysis}

A key objective of observing multiple active apparitions of MBCs is to be able to study the evolution of activity over time for these objects.
In order to investigate the evolution of 259P's activity, we focus on data from 2017 September 17 to 2017 December 22, which span a true anomaly range of $20.5^{\circ}<\nu<59.7^{\circ}$.  These observations cover a similar orbital arc as the 2008 data reported by \citet{jewitt2009_259p}, which span a true anomaly range of $29.2^{\circ}<\nu<48.6^{\circ}$.  While 259P had another perihelion passage (on 2013 January 25) between its 2008 and 2017 active apparitions, it unfortunately was not observable from the Earth until well after that perihelion passage, and as such, we have no observational coverage of any activity around that time period that can be included in this analysis.

\begin{figure}[tbp]
\centerline{\includegraphics[width=3.3in]{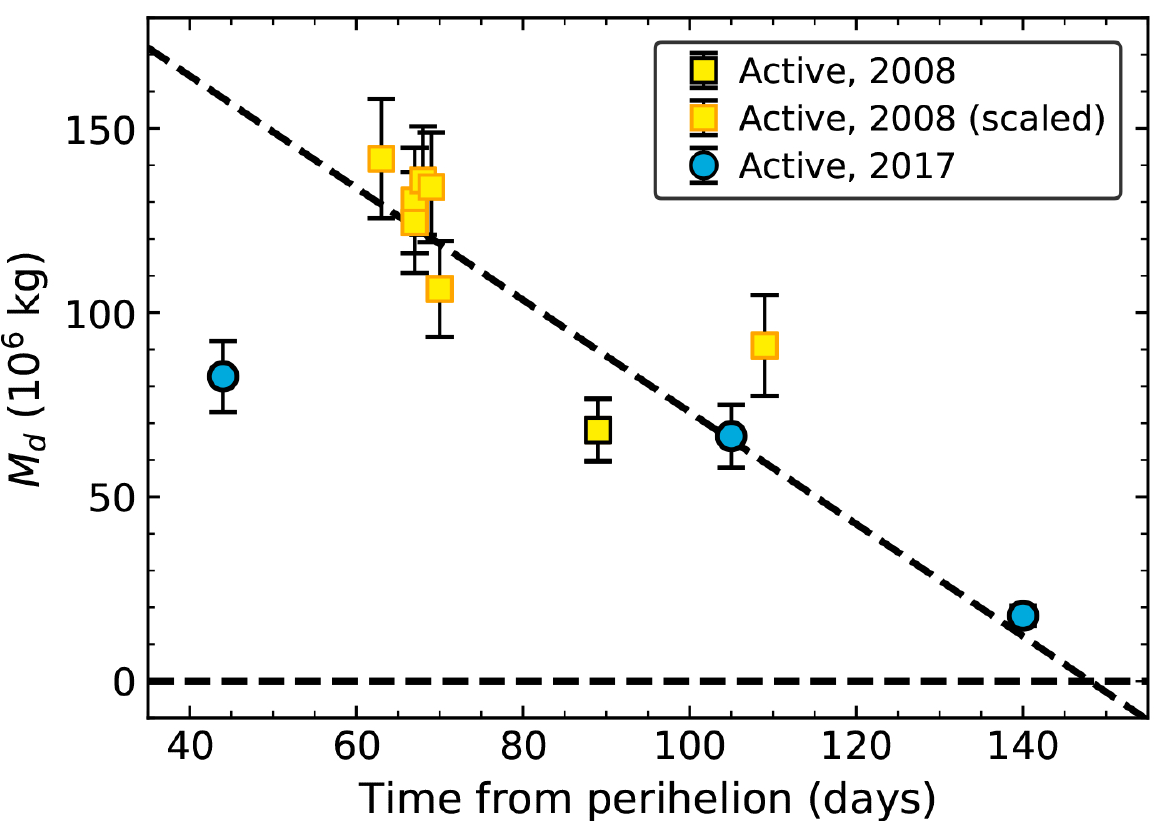}}
\caption{\small Estimated total dust masses for 259P during observations from 2008 and 2017 plotted versus time from perihelion (where positive values denote time after perihelion).  A diagonal dashed line shows a linear fit to data from 2008.  Dust masses computed from total fluxes measured as part of this work using archival data from 2008 and newly reported data from 2017 are marked with a yellow-filled square and blue-filled circles with black outlines, respectively, while dust masses estimated from previously reported photometry as described in Section~\ref{subsection:photometric_analysis} are marked with yellow-filled squares with orange outlines.}
\label{figure:fading}
\end{figure}

Re-plotting the dust masses inferred from photometry measurements of our 2017 observations (Table~\ref{table:obs_259p}) and total dust masses estimated from previously reported photometry (Table~\ref{table:obs_259p_previous}) as a function of time after perihelion (Figure~\ref{figure:fading}), we find that dust mass estimates from data from 2017 November 17 \revision{(obtained when 259P was 105 days past perihelion)} and 2017 December 22 \revision{(140 days past perihelion)} are consistent with a linear fading function fit to scaled dust mass estimates for the 2008 observations of 259P.  We also see from Figure~\ref{figure:fading} that the total dust mass estimated for the comet on 2017 September 17 (44 days past perihelion) is substantially smaller than the estimated total dust mass inferred to be present at the start of the 2008 observations.  However, if the visible dust mass in 2017 continued \revision{its then upward trend past}
the 2017 September 17 observations, it could \revision{reasonably} conceivably have reached a level comparable to that inferred to be present at the start of the 2008 observations (63 days past perihelion) before then declining to its observed level on 2017 November 17.  We conclude from this analysis that there is no clear evidence of changes in 259P's activity strength between its 2008 and 2017 active apparitions.


\section{DISCUSSION}\label{section:discussion}

Multiple active apparitions have now been observed and characterized for seven MBCs \citep[133P/Elst-Pizarro, 238P/Read, 259P, 288P/(300163) 2006 VW$_{139}$, 313P/Gibbs, 324P/La Sagra, and 358P/PANSTARRS;][and this work]{hsieh2004_133p,hsieh2011_238p,hsieh2015_313p,hsieh2018_358p,hsieh2015_324p,agarwal2016_288p_cbet},
although direct comparisons of activity strength measured \revision{using methods similar to those performed in this work} over similar orbit arcs has only been performed thus far for 238P, 259P, and 288P.  
Within this group of objects, 259P is the only one with a semimajor axis interior to the 5:2 mean-motion resonance with Jupiter, and a perihelion distance of $q<2$~au.
While we find 259P's activity strength to remain roughly similar between 2008 and 2017 in this work, \citet{hsieh2018_238p288p} found that 238P's activity appeared to decrease between its 2010-2011 active apparition and its 2016-2017 active apparition, while 288P's activity appeared to increase between its 2000 active apparition and its 2016 active apparition.  In the cases of both 238P and 288P, activity onset times were determined to have remained similar for each active apparition, suggesting that only minimal changes in ice depth (e.g., due to sublimation-driven ice recession or mantling) had occurred between apparitions \citep{hsieh2018_238p288p}.
An intensive observing campaign was conducted in 2011 for an eighth MBC, 176P/LINEAR, during the perihelion passage following its first observed active apparition in 2005, but while those 2011 observations overlapped observations of the object's 2005 active apparition in terms of orbit positions, no evidence of activity was found \citep{hsieh2014_176p}, suggesting that a decline of detectable activity strength to zero had occurred from the first epoch to the next.

\setlength{\tabcolsep}{4.0pt}
\setlength{\extrarowheight}{0em}
\begin{table*}[htb!]
\caption{Activity parameters measured for MBCs$^a$}
\centering
\smallskip
\footnotesize
\begin{tabular}{lccccccccccc}
\hline\hline
\multicolumn{1}{c}{Object$^b$}
 & \multicolumn{1}{c}{$a^c$}
 & \multicolumn{1}{c}{$q^d$}
 & \multicolumn{1}{c}{$r_N^e$}
 & \multicolumn{1}{c}{$\nu_0^f$}
 & \multicolumn{1}{c}{$r_{h,0}^g$}
 & \multicolumn{1}{c}{$T_0^h$}
 & \multicolumn{1}{c}{$\dot M_d^i$}
 & \multicolumn{1}{c}{$r_{\dot M}$$^j$}
 & \multicolumn{1}{c}{$A_{\rm act}^k$}
 & \multicolumn{1}{c}{$f_{\rm act}^l$}
 & \multicolumn{1}{c}{Ref.$^m$}
 \\
\hline
238P (2010) & 3.162 & 2.361 & 0.4 & 302$\pm$12 & 2.61$\mp$0.11 & 171$-$190 & 1.4$\pm$0.3 & 2.49 & 6$\times$10$^{3}$ $-$ 1$\times$10$^{5}$ & 3$\times$10$^{-3}$ $-$ 7$\times$10$^{-2}$ & [1] \\
238P (2016) & 3.162 & 2.361 & 0.4 & 297$\pm$21 & 2.66$\mp$0.19 & 170$-$190 & 0.7$\pm$0.3 & 2.39 & 2$\times$10$^3$ $-$ 4$\times$10$^4$ & 1$\times$10$^{-3}$ $-$ 2$\times$10$^{-2}$ & [1] \\
259P (2017) & 2.727 & 1.796 & 0.30 & 313.9$\pm$0.4 & 1.96$\mp$0.03 & 182$-$195 & \revision{4.6$\pm$0.3} & 1.86 & \revision{8$\times$10$^3$ $-$ 7$\times$10$^4$} & \revision{7$\times$10$^{-3}$ $-$ 6$\times$10$^{-2}$} & [2] \\
288P (2010) & 3.047 & 2.434 &  2$\times$0.80 & 332$\pm$4 & 2.51$\mp$0.02 & 173$-$191 & 3.5$\pm$0.4 & 2.47 & 1$\times$10$^4$ $-$ 3$\times$10$^5$ & 9$\times$10$^{-4}$ $-$ 2$\times$10$^{-2}$ & [1] \\
288P (2016) & 3.047 & 2.434 & 2$\times$0.80 & 333$\pm$4 & 2.48$\mp$0.02 & 174$-$191 & 5.6$\pm$0.7 & 2.44 & 2$\times$10$^4$ $-$ 5$\times$10$^5$ & 1$\times$10$^{-3}$ $-$ 3$\times$10$^{-2}$ & [1] \\
324P (2015) & 3.096 & 2.620 & 0.55 & $<$300 & $>$2.80 & $<$189 & $<$0.1 & 2.78 & $<$2$\times$10$^5$ & $<$4$\times$10$^{-2}$ & [3] \\
358P (2017) & 3.155 & 2.410 & 0.32 & 316$\pm$1 & 2.54$\mp$0.01 & 172$-$190 & 2.0$\pm$0.6 & 2.50 & 4$\times$10$^4$ $-$ 1$\times$10$^6$ & 3$\times$10$^{-2}$ $-$ 8$\times$10$^{-1}$ & [4] \\
\hline
\hline
\multicolumn{12}{l}{\revision{$^a$ Activity parameters measured for MBCs using the methods described in Section~\ref{subsection:photometric_analysis}.}} \\
\multicolumn{12}{l}{\revision{$^b$ Object designations with the perihelion year corresponding to the analyzed apparition in parentheses.}} \\
\multicolumn{12}{l}{$^c$ Semimajor axis, in au.} \\
\multicolumn{12}{l}{$^d$ Perihelion distance, in au.} \\
\multicolumn{12}{l}{$^e$ Nucleus radius, in km.} \\
\multicolumn{12}{l}{$^f$ True anomaly, in degrees, at time of \revision{estimated} activity onset.} \\
\multicolumn{12}{l}{$^g$ Heliocentric distance, in au, at time of \revision{estimated} activity onset.} \\
\multicolumn{12}{l}{$^h$ Estimated equilibrium surface temperature, in K, at time of \revision{estimated} activity onset.} \\
\multicolumn{12}{l}{$^i$ Estimated \revision{average} net dust production rate early in activity period, in kg~s$^{-1}$.} \\
\multicolumn{12}{l}{$^j$ Heliocentric distance, in au, at the midpoint of the time period to which the indicated \revision{average} dust production rate corresponds.} \\
\multicolumn{12}{l}{$^k$ Estimated effective active surface area, in m$^2$, corresponding to the indicated \revision{average} dust production rate.} \\
\multicolumn{12}{l}{$^l$ Estimated effective active fractional surface area corresponding to the indicated \revision{average} dust production rate.} \\
\multicolumn{12}{l}{$^m$ References:
[1] \citet{hsieh2018_238p288p}; 
[2] \citet{maclennan2012_259p},
    This work;
[3] \citet{hsieh2014_324p},
    \citet{hsieh2015_324p};} \\
\multicolumn{12}{l}{~~~~~[4] \citet{hsieh2018_358p}} \\
\end{tabular}
\label{table:mbc_comparisons}
\end{table*}

\setlength{\tabcolsep}{4.0pt}
\setlength{\extrarowheight}{0em}

The diversity of activity evolution behavior observed for just four objects suggests that there could be competing processes modulating activity strength on MBCs.  For example, while volatile depletion and mantling would be expected to cause overall declines in activity strength over time \citep[e.g.,][]{kossacki2012_259p,hsieh2015_ps1mbcs}, processes such as sinkhole collapses and expansion of active areas via sublimation-driven erosion, as observed on 67P/Churyumov-Gerasimenko by the Rosetta spacecraft \citep[e.g.,][]{vincent2015_cometsinkholes}, could conceivably produce occasional increases in activity strength.  If active sites are localized \citep[e.g.,][]{hsieh2004_133p,yu2020_elstpizarro}, variable global or local seasonal effects due to rotational axis precession or topographical surface evolution between active apparitions \citep[e.g.,][]{gutierrez2016_67Pprecession,elmaarry2019_cometevolution,lai2019_67pseasonaleffects,marschall2020_cometcomasurfacelinks} could also conceivably alter MBC activity strength from orbit to orbit.
Continued investigation of activity evolution in MBCs to ascertain the relative frequencies of different evolutionary trends will be extremely valuable for illuminating the significance of different processes in driving the evolution of MBC surfaces.


The heliocentric distance of 259P's activity onset point ($r_{h,0}=1.96$~au) is significantly smaller than those of other MBCs, whose estimated activity onset points measured thus far have consistently had heliocentric distances of $r_{h,0}\sim2.5-2.6$~au (Table~\ref{table:mbc_comparisons}).  Once activated, however, 259P's \revision{average} net \revision{pre-perihelion} dust production rate appears to be comparable to those of other MBCs (Table~\ref{table:mbc_comparisons}).  The much smaller heliocentric distance of 259P's activity onset point relative to other MBCs suggest that its ice reservoirs may be located at greater depths than on MBCs farther from the Sun, increasing the time needed for a \revision{solar irradiation-driven} thermal wave to reach subsurface ice, even after the object has reached a heliocentric distance at which sublimation-driven activity on other MBCs would normally begin (which for 259P would correspond to $\nu\sim260-265^{\circ}$ as it approaches perihelion, rather than $\nu_0\sim315^{\circ}$ at which its activity is actually estimated to begin).  Deeper ice on 259P could be a result of more rapid ice depletion caused by the object's closer proximity to the Sun (and therefore greater total solar flux and higher surface temperatures) compared to other MBCs.
Alternatively, 259P's activity could instead be primarily modulated by other mechanisms, such as seasonal effects \citep[e.g.,][]{bertini2012_81p,li2016_c2013a1,eisner2017_49p}.
Detailed characterization of the activity of other MBCs in the middle main belt (i.e., with semimajor axes between the 3:1 and 5:2 mean-motion resonances with Jupiter at $a=2.502$~au and $a=2.824$~au, respectively), such as P/2015 X6 ($a=2.755$~au; $q=2.287$~au), will provide an important point of comparison for assessing how typical the activity characteristics we measure for 259P are of this MBC sub-group.

\section{Conclusions}\label{section:conclusions}

In this work, we present analyses of observations using the Gemini North and South telescopes prior to and following MBC 259P/Garradd's 2017 perihelion passage showing its reactivation \citep[previously reported by][]{hsieh2017_259p}.  We report the following key results:
\begin{itemize}
    \item{We confirm the reactivation of 259P, previously reported by \citet{hsieh2017_259p}, and present monitoring data following the evolution of the object's activity for nine months spanning its 2017 perihelion passage on UT 2017 August 4, covering a true anomaly range of $\nu=-55.6^{\circ}$ to $\nu=59.7^{\circ}$. These observations place 259P among seven MBCs in total that have been confirmed to have recurrent activity, where 259P is the only one among those with a semimajor axis interior to the 5:2 mean-motion resonance with Jupiter, and a perihelion distance of $q<2$~au.}
    \item{Using an estimated mean effective dust grain radius of \revision{${\bar a}_d\sim2$~mm} \revision{and assuming a dust grain density of $\rho=2500$~kg~m$^{-3}$}, we find a best-fit \revision{average} net \revision{pre-perihelion} dust production rate for 259P during its 2017 active period of \revision{$\dot M_d=(4.6\pm0.2)$~kg~s$^{-1}$} and a best-fit start date of activity of 2017 April $22\pm1$ ($\sim104$~days prior to perihelion), when 259P was at \revision{$r_h=(1.96\mp0.03)$~au} and \revision{$\nu=(313.9\pm0.4)^{\circ}$}.  \revision{From this calculated average dust production rate,} we estimate the effective active fraction of the object's surface to be from \revision{$f_{\rm act}\sim7\times10^{-3}$} to \revision{$f_{\rm act}\sim6\times10^{-2}$} (corresponding to effective active areas of \revision{$A_{\rm act}\sim8\times10^3$~m$^2$} to \revision{$A_{\rm act}\sim7\times10^4$~m$^2$}) at the start of its 2017 active period.}
    \item{A comparison of estimated total dust masses measured for 259P in 2008 and 2017 shows no clear evidence of changes in the object's activity strength between those two active apparitions.}
\end{itemize}

\section*{Acknowledgments}\label{section:acknowledgments}
We are grateful to two anonymous reviewers who helped to improve this manuscript.  HHH, MMK, NAM, and SSS acknowledge support from the NASA Solar System Observations program (Grants NNX16AD68G and 80NSSC19K0869).  We are grateful to J.\ Chavez, J.\ Fuentes, M.\ Gomez, A.\ Lopez, L.\ Magill, E.\ Marin, K.\ Roth, D.\ Sanmartim, A.\ Shugart, K.\ Silva, A.\ Smith, and E.\ Wenderoth for assistance in obtaining observations.  This research made use of {\tt astropy}, a community-developed core {\tt python} package for astronomy, and {\tt uncertainties} (version 3.0.2), a {\tt python} package for calculations with uncertainties by E.\ O.\ Lebigot. This work benefited from support by the International Space Science Institute in Bern, Switzerland, through the hosting and provision of financial support for an international team, which was led by C.\ Snodgrass and included HHH and MMK, to discuss the science of MBCs.  This work is based on observations obtained at the Gemini Observatory (Program IDs GN-2013A-Q-102, GS-2017A-LP-11, and GS-2017B-LP-11), which is operated by the Association of Universities for Research in Astronomy, Inc., under a cooperative agreement with the NSF on behalf of the Gemini partnership: the National Science Foundation (United States), the National Research Council (Canada), CONICYT (Chile), Ministerio de Ciencia, Tecnolog\'{i}a e Innovaci\'{o}n Productiva (Argentina), and Minist\'{e}rio da Ci\^{e}ncia, Tecnologia e Inova\c{c}\~{a}o (Brazil).  This work is also based in part on data collected at Subaru Telescope and obtained from the SMOKA, which is operated by the Astronomy Data Center, National Astronomical Observatory of Japan.  We wish to recognize and acknowledge the very significant cultural role and reverence that the summit of Maunakea has always had within the indigenous Hawaiian community.  We are fortunate to have the opportunity to conduct observations from this mountain.

%

\facilities{Gemini:North (GMOS), Gemini:South (GMOS), Subaru (FOCAS)}


\software{{\tt astropy} \citep{astropy2018_astropy},
    {\tt astroquery} \citep{ginsburg2019_astroquery},
    {\tt IRAF} \citep{tody1986_iraf,tody1993_iraf}, 
    {\tt L.A.Cosmic} \citep{vandokkum2001_lacosmic,vandokkum2012_lacosmic},
    {\tt uncertainties} (v3.0.2, E.\ O.\ Lebigot),
    {\tt refcat} \citep{tonry2018_refcat}
          }





\bibliography{hhsieh_refs}{}
\bibliographystyle{aasjournal}



\end{document}